# Multiplication and division of the orbital angular momentum of light with diffractive transformation optics

Gianluca Ruffato[1,2], Michele Massari[2,3], and Filippo Romanato[1,2,3]

[1]Department of Physics and Astronomy 'G. Galilei', University of Padova, via Marzolo 8, 35131 Padova, Italy

[2]LaNN, Laboratory for Nanofabrication of Nanodevices, EcamRicert, Corso Stati Uniti 4, 35127 Padova, Italy.

[3]CNR-INFM TASC IOM National Laboratory, S.S. 14 Km 163.5, 34012 Basovizza, Trieste, Italy

Authors e-mails:

Gianluca Ruffato: gianluca.ruffato@unipd.it

Michele Massari: massari@iom.cnr.it

Filippo Romanato: filippo.romanato@unipd.it

Correspondence: Gianluca Ruffato, via Marzolo 8, Department of Physics and Astronomy 'G. Galilei', University of Padova, 35125 Padova (Italy), tel. +390498277081, fax. +390498277102

# ABSTRACT

We present a method to efficiently multiply or divide the orbital angular momentum (OAM) of light beams using a sequence of two optical elements. The key-element is represented by an optical transformation mapping the azimuthal phase gradient of the input OAM beam onto a circular sector. By combining multiple circular-sector transformations into a single optical element, it is possible to perform the multiplication of the value of the input OAM state by splitting and mapping the phase onto complementary circular sectors. Conversely, by combining multiple inverse transformations, the division of the initial OAM value is achievable, by mapping distinct complementary circular sectors of the input beam into an equal number of circular phase gradients. The optical elements have been fabricated in the form of phase-only diffractive optics with high-resolution electron-beam lithography. Optical tests confirm the capability of the multiplier optics to perform integer multiplication of the input OAM, while the designed dividers are demonstrated to correctly split up the input beam into a complementary set of OAM beams. These elements can find applications for the multiplicative generation of higher-order OAM modes, optical information processing based on OAM-beams transmission, and optical routing/switching in telecom.

# INTRODUCTION

Since the seminal paper of Allen and coworkers [1], light beams carrying orbital angular momentum (OAM) gave rise to a flourishing field of research, leading to a rich multiplicity of studies and applications [2]: particle trapping, tweezing and manipulation [3], high-resolution microscopy [4, 5], astronomical coronagraphy [6], mode-division multiplexing [7] and security [8, 9]. Concurrently, several optical methods and devices have been described and engineered in order to tailor and control the OAM content of light beams, for instance spiral phase plates [10, 11], computer-generated holograms [12, 13], *q*-plates [14], transformation optics [15, 16] and metasurfaces [17]. Basically, all those methods rely on transferring an azimuthal phase gradient $exp(i\ell\varphi)$ to the input beam, being $\varphi$ the polar angle on a plane transverse to the propagation direction, and $\ell$ the amount of OAM per photon in units of $\hbar$. In microscopy, a beam carrying OAM was used for suppressing fluorescence in the dark centre zone providing sub-wavelength lateral confinement and consequent super-resolution as shown by the pioneering paper of Hell *et al.* [18]. In the telecom field, the potentially unbounded state space provided by this even-unexploited degree of freedom opens to a promising solution to increase the information capacity of optical networks [19], both for free-space [20] and optical fiber propagation [21]. Nowadays, it becomes urgent to further develop novel devices that can reconfigure and switch between distinct OAM modes dynamically [22, 23] in order to fully exploit the extra degree of freedom provided by the OAM for both classical and quantum communications. The above-mentioned conventional methods are useful to implement only shift operations on the OAM mode, i.e. addition or subtraction. On the other hand, it would be extremely beneficial to be able to either multiply or divide the OAM state for some applications, such as the multiplicative generation of high-order modes, optical switching and routing [24, 25], and optical OAM-based information processing [26, 27].

So far, OAM multiplication and division have been implemented by means of bulky and sophisticated solutions [28-30] which are barely suitable to integration and miniaturization, due to the presence of many optical elements. Multiplication has been performed [28, 29] by mapping the input azimuthal phase gradient into a linear phase gradient with a first *log-pol* sorter [31, 32], then creating multiple copies with an optical fan-out [33], and finally wrapping the extended phase gradient into a doughnut shape with a second *log-pol* sorter, working in reverse, for a total of at least 6 optical elements, plus lenses in between for Fourier transform and beam resizing. Division has been implemented in a similar manner [29, 30], by substituting the multiple-copy stage with a mask for the selection of only a fraction of the linear phase gradient in input, therefore at the expense of a non-negligible amount of the input energy. In the quest for miniaturization, the implementation of the *log-pol* transformation stages in the form of diffractive optics [34], and the integration of the copying/fractioning operation in the *log-pol* optics, [35] could improve the compactness of the system. However, the number of optical operations, i.e. *log-pol* mapping, phase-resizing and inverse transformation, would remain the same.

Here we present a completely different method, which basically preserves the axial symmetry and avoids the limitations of the *log-pol* coordinate-change approach. The key-element is represented by an optical transformation which maps the azimuthal phase gradient of the input OAM beams into a circular sector. By combining multiple circular-sector transformation onto a single optical element, it is possible to perform the multiplication of the input OAM by mapping its phase onto complementary circular sectors. Conversely, by combining multiple inverse transformations, it is possible to map different complementary sectors of the input beam into an equal number of circular phase gradients, thus achieving a division of the initial OAM. These operations can be realized by a sequence of only two elements, performing the optical transformation of the beam and the required phase-correction, respectively. This approach allows to perform multiplication and division of OAM in a compact manner, remarkably reducing the number of optical operations and the total amount of optical elements, and therefore providing a final significant increase in the optical efficiency. The designed optical elements have been fabricated in the form of phase-only diffractive optics with high-resolution electron-beam lithography, and optically characterized in order to demonstrate the expected capability to either multiply or divide the OAM of the input beam. In addition, a compact optical architecture has been presented in order to arrange the two optical elements, required for optical transformation and phase correction, onto the same substrate, further improving alignment and miniaturization.

## MATERIALS AND METHOS

### SIMULATON AND DESIGN

In the following paragraphs, the theory underlying multiplication and division with transformation optics is presented and described. At first, the key element of these operations, i.e. the circular-sector transformation, is introduced and developed in the paraxial approximation. Therefore, the possibility to combine many circular-sector transformations in order to either multiply or divide the OAM content of an input beam is shown, and the phase patterns of the corresponding optical elements are calculated. Numerical simulations have been performed with MatLab, implementing the convolution algorithm in the paraxial regime [36].

*Circular-sector transformation*

The optical layout of the system is constituted of a cascade of two optical elements: the former performs a conformal optical transformation, while the latter corrects the phase distortion due to the different paths travelled by the distinct points of the beam and restores the desired phase profile. Conversely, due to the invariance of the light path for time reversal, the second optical element performs the inverse optical transformation. The key-element of OAM multiplication and division is represented by an optical transformation performing a conformal mapping of the whole circle onto a circular sector (Fig. 1a and 1d). By indicating with ($r$, $\vartheta$) the polar coordinates on the input plane, and with ($\rho$, $\phi$) the polar reference frame on the second plane, such transformation operates a rescaling of the azimuthal coordinate:

$$\varphi = \frac{\vartheta}{n} \quad (1)$$

After imposing the condition of conformity, the new radial coordinate is given by:

$$\rho = a\left(\frac{r}{b}\right)^{-\frac{1}{n}} \quad (2)$$

being *a* and *b* scaling parameters. By applying the previous transformation, an azimuthal phase gradient is mapped conformally onto a circular sector with amplitude $2\pi/n$. In order to calculate the phase pattern of an optical element performing this transformation in the paraxial regime, we apply the stationary phase approximation [37] to the Fresnel integral. The field $U(\rho,\varphi,f)$ after a propagation length *f* for an input plane-wave illuminating a phase-only optical element with phase function $\Omega_S$, located at $z=0$, is given by:

$$U(\rho,\varphi) = \frac{e^{ik\frac{\rho^2}{2f}}}{i\lambda f} \iint e^{i\Omega_{S,n}(r,\vartheta)} e^{ik\frac{r^2}{2f}} e^{-ik\frac{r\rho}{f}\cos(\vartheta-\varphi)} r\,dr\,d\vartheta \quad (3)$$

According to the stationary phase approximation [37], the integral solution reduces to find the saddle points of the phase function:

$$\Phi(r,\vartheta) = \Omega_{S,n}(r,\vartheta) + k\frac{r^2}{2f} - k\frac{r\rho}{f}\cos(\vartheta-\varphi) \quad (4)$$

The condition $\nabla\Phi = 0$ leads to a system of partial derivatives of $\Omega_S$ unknown. Substituting the transformation relations, i.e. eqs. (1) and (2), we get:

$$\begin{cases} \dfrac{\partial \Omega_{S,n}}{\partial r} + k\dfrac{r}{f} - k\dfrac{a}{f}\left(\dfrac{r}{b}\right)^{-\frac{1}{n}} \cos\left(\vartheta - \dfrac{\vartheta}{n}\right) = 0 \\ \dfrac{\partial \Omega_{S,n}}{\partial \vartheta} + k\dfrac{ar}{f}\left(\dfrac{r}{b}\right)^{-\frac{1}{n}} \sin\left(\vartheta - \dfrac{\vartheta}{n}\right) = 0 \end{cases} \qquad (5)$$

After integrating, we obtain the analytical expression for the phase function in the paraxial regime:

$$\Omega_{S,n}(r,\vartheta) = \frac{2\pi ab}{\lambda f}\left(\frac{r}{b}\right)^{1-\frac{1}{n}} \cdot \frac{\cos\left[\vartheta\left(1-\dfrac{1}{n}\right)\right]}{1-\dfrac{1}{n}} - k\frac{r^2}{2f} \qquad (6)$$

The corresponding phase-corrector can be calculated by applying the same method to the inverse transformation. We obtain:

$$\Omega_{PC,n}(\rho,\varphi) = \frac{2\pi ab}{\lambda f}\left(\frac{\rho}{a}\right)^{1-n} \cdot \frac{\cos\left[\varphi(1-n)\right]}{1-n} - k\frac{\rho^2}{2f} \qquad (7)$$

which is basically the expression in eq. (6) after the substitutions $b \to a$, $n \to 1/n$.

In Fig. 1a and Fig. 1d, schematics of OAM-beams evolution in case of circular-sector transformations by a factor of $n=2$ and $n=3$ are depicted, respectively. The phase gradient of the input OAM beam is mapped over half and one-third of the entire circle. Numerical simulations are reported (Fig. 1c, Fig. 1f), showing the evolution of the beam after illuminating the first optical element and during propagation in the range of the focal length, where the second element is placed for phase-correction.

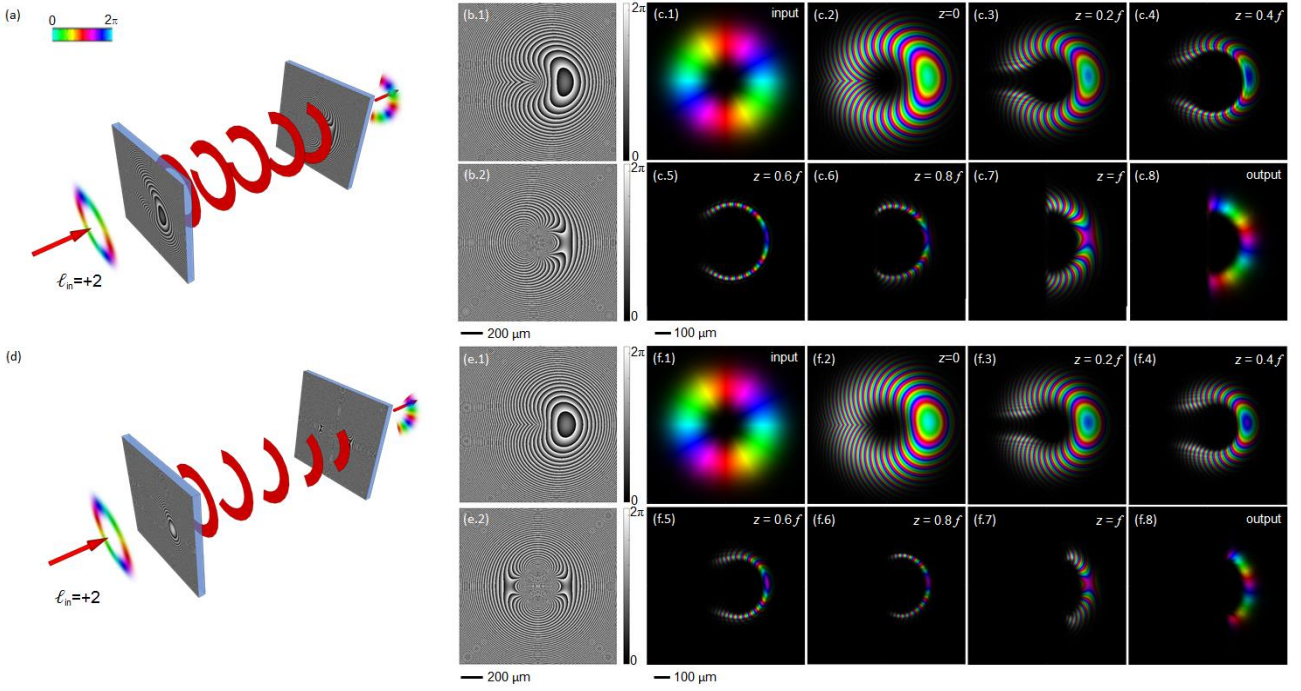

**Figure 1**. Schematics of OAM-beams transformation with circular-transformation optics, for two-fold ($n=2$) (a) and three-fold ($n=3$) (d) circular-sector transformation. The first phase-pattern (b.1, e.1) performs an $n$-fold circular-sector transformation, mapping the input intensity distribution onto a $2\pi/n$ arc. The second phase-pattern (b.2, e.2) performs the required phase-correction, retaining the compressed azimuthal-phase distribution. Design parameters: $a=300$ μm, $b=250$ μm, $f=20$ mm. Numerical simulations of the propagation of an input Laguerre-Gaussian beam carrying $\ell=2$ (c.1, f.1), after illuminating the first element (c.2, f.2), up to the second optical element (c.7, f.7), and output phase-corrected beam (c.8, f.8). Colours and brightness refer to phase and intensity, respectively.

*The OAM multiplier*

OAM multiplication by a factor $n$ basically consists in splitting the input beam into $n$ copies and mapping each one over a set of $n$ complementary circular sectors. Therefore, the phase pattern $\Omega_{M,n}$ of the multiplier is described by the superposition of $n$ circular-sector transformations $\{\Omega_{S,n}^{(j)}\}$, $j=1,\ldots n$, mapping the input beam over the corresponding circular sectors with amplitude $2\pi/n$ and centered in $\{(j-1)2\pi/n\}$:

$$\Omega_{M,n}(r,\vartheta) = \arg\left\{\sum_{j=1}^{n} e^{i\Omega_{S,n}^{(j)}}\right\} \tag{8}$$

where:

$$\Omega_{S,n}^{(j)}(r,\vartheta) = \frac{2\pi ab}{\lambda f}\left(\frac{r}{b}\right)^{1-\frac{1}{n}} \cdot \frac{\cos\left[\vartheta\left(1-\frac{1}{n}\right)+(j-1)\frac{2\pi}{n}\right]}{1-\frac{1}{n}} - k\frac{r^2}{2f} \tag{9}$$

and the corresponding phase-corrector is given again by eq. (7).

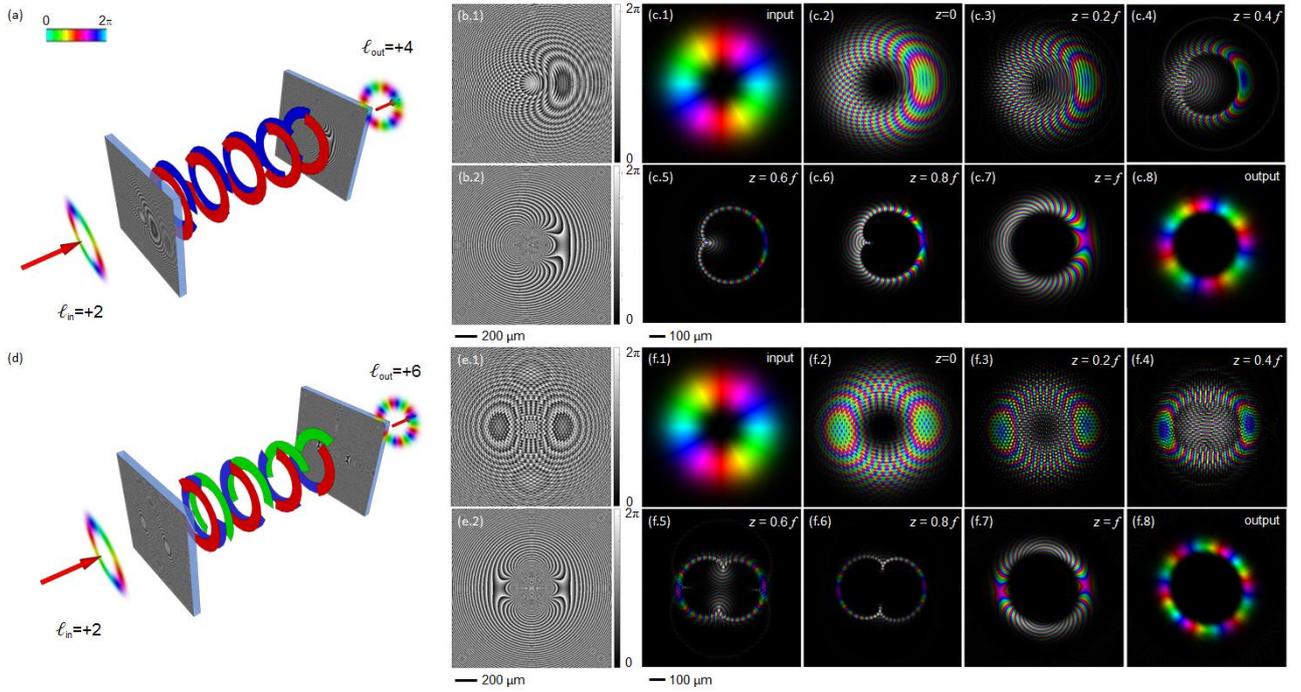

**Figure 2**. Schematics of OAM-beams optical multipliers for two-fold ($n=2$) (a) and three-fold ($n=3$) (d) multiplication with transformation optics. The first phase-pattern (b.1, e.1) performs an $n$-fold multiplication, splitting and mapping the input azimuthal phase gradient over $n$ complementary arcs. The second phase-pattern (b.2, e.2) performs phase-correction, retaining the azimuthal phase distribution. Design parameters: $a=300$ μm, $b=250$ μm, $f=20$ mm. Numerical simulations of the propagation of an input Laguerre-Gaussian beam carrying $\ell=2$ (c.1, f.1), after illuminating the first element (c.2, f.2), up to the second optical element (c.7, f.7), and output phase-corrected beam (c.8, f.8). Colours and brightness refer to phase and intensity, respectively.

In Fig. 2a and Fig. 2d, schematics of the beam evolution in case of multiplication by a factor of $n=2$ and $n=3$ are depicted, respectively. In Fig. 2c and 2f, the corresponding numerical simulations are reported, providing the evolution of the beam in the range of the focal length, where the second element is placed for phase-correction. As expected, the input beam is split into $n$ equal contributions which are mapped onto complementary arcs, thus performing an $n$-fold multiplication of the input OAM.

*The OAM divider*

It could be interesting to divide the input beam into a bunch of $n$ beams with OAM equal to $1/n$ the OAM of the input beam, as depicted in Fig. 3a and Fig. 3d, for the case of two-fold ($n=2$) and three-fold ($n=3$) division, respectively. This is possible by defining the phase pattern of the OAM divider in the following way:

$$\Omega_{D,n}(r,\vartheta) = \arg\left\{\sum_{j=1}^{n} e^{i\Omega_{S,1/n}^{(j)}(r,\vartheta+(j-1)2\pi/n)} e^{ir\cdot\beta^{(j)}\cos(\vartheta-\vartheta^{(j)})} \Theta\left(\vartheta - \frac{2\pi}{n}(j-1) + \frac{\pi}{n}\right) \Theta\left(\frac{2\pi}{n}(j-1) + \frac{\pi}{n} - \vartheta\right)\right\} \quad (10)$$

being $\Theta(\cdot)$ the Heaviside function ($\Theta(x)=1$, when $x\geq 0$, $\Theta(x)=0$ otherwise). According to eq. (10), the total phase pattern of the OAM divider is basically the composition of $n$ complementary and non-overlapping phase-patterns, each one defined over a circular sector spanning an angle equal to $2\pi/n$. In the specific, each zone is illuminated by $1/n$th of the input beam and it is designed in order to wrap the impinging arc onto the entire circle ($1/n$-fold circular-sector transformation). In addition, spatial frequency carriers $\{(\beta^{(j)}, \vartheta^{(j)})\}$, are introduced, in order to spatially separate the $n$ output beams and locate them at defined positions ($\rho^{(j)}, \phi^{(j)}$) in the focal plane, according to:

$$\begin{cases} \rho^{(j)} = \frac{f}{k}\beta^{(j)} \\ \phi^{(j)} = \vartheta^{(j)} = (j-1)\frac{2\pi}{n} \end{cases} \quad (11)$$

where $j=1,\ldots n$, being $f$ the focal length of the optical transformation. The phase-corrector is given by the composition of $n$ complementary and non-overlapping phase-correctors, each one defined over a circular sector spanning an angle equal to $2\pi/n$ and centered at the positions given by $(\rho^{(j)}, \phi^{(j)})$:

$$\Omega_{D-PC,n}(\rho,\varphi) = \arg\left\{\sum_{j=1}^{n} e^{i\Omega_{PC,1/n}^{(j)}\left(\rho'_j,\varphi'_j+(j-1)2\pi/n\right)}\Theta\left(\varphi - \frac{2\pi}{n}(j-1) + \frac{\pi}{n}\right)\Theta\left(\frac{2\pi}{n}(j-1) + \frac{\pi}{n} - \varphi\right)\right\} \quad (12)$$

where $(\rho'_j, \varphi'_j)$ are polar coordinates centered in $(\rho^{(j)}, \phi^{(j)})$. In Figures 3c and 3f, numerical calculations are reported in the case of $n=2$ and $n=3$, providing the evolution of the beam in the range of the focal length. As expected, the input beam is split into $n$ equal contributions which are mapped onto distinct circles, thus performing an $n$-fold division of the input OAM into a set of $n$ OAM beams. In principle, the OAM divider can be generalized and properly designed in order to decompose the input OAM into many beams carrying different values of OAM and whose total sum equals the OAM value of the input beam.

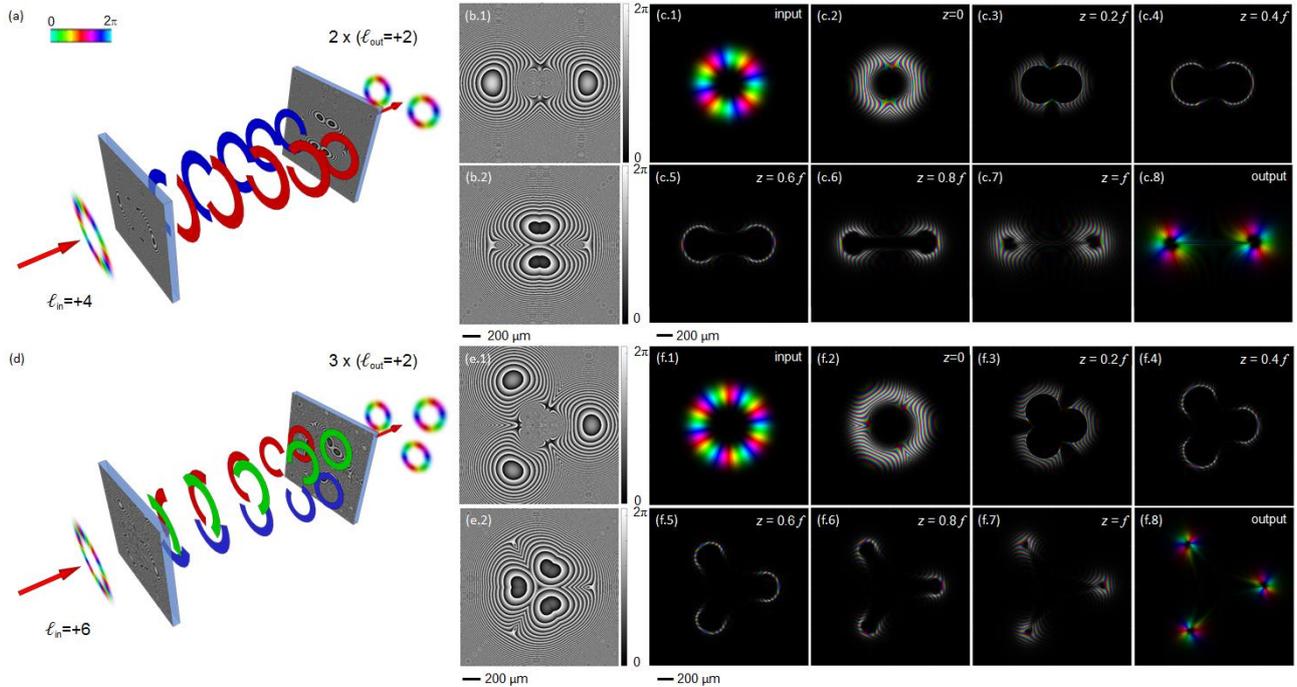

**Figure 3**. Schematics of OAM-beams optical dividers, for two-fold ($n=2$) (a) and three-fold ($n=3$) (d) division with transformation optics. The first phase-pattern (b.1, e.1) performs an $n$-fold division, splitting

the input azimuthal gradient into *n* complementary arcs which are wrapped and mapped over distinct whole circles at desired positions. The second phase-pattern (b.2, e.2) performs phase-correction, retaining the azimuthal phase distribution for each output beam. Design parameters: $a$=300 μm, $b$=300 μm, $f$=20 mm, spatial frequency $\beta^{(j)}$= 0.3 μm$^{-1}$, $j$=1,..,3. Numerical simulations of the propagation of an input Laguerre-Gaussian beam carrying $\ell$=4 (c.1) or $\ell$=6 (c.1), after illuminating the first element (c.2, f.2), up to the second optical element (c.7, f.7), and output phase-corrected beams (c.8, f.8). Colours and brightness refer to phase and intensity, respectively.

## *Optical elements design*

A limit of performing optical operations with transformation optics is the need for at least two optical elements, i.e. transformer and phase-corrector. As widely experienced in the case of OAM sorting with *log-pol* transformation, the presence of two confocal elements to be perfectly aligned, coaxial and coplanar, could require arduous efforts for alignment, and could result detrimental for miniaturization and integration. In order to simplify the alignment process and improve the compactness of the optical architecture, we designed the optical configuration in order to incorporate the two elements onto a single optical platform. In this configuration, the substrate is illuminated twice: after crossing the first zone, performing multiplication/division, the beam is back-reflected with a mirror, placed at half the focal length, and impinges on the second region providing the phase-corrector (Fig. 6b). By adding a tilt term to the first phase pattern, the back-reflected beam does not overlap the input one and the two optical elements, i.e. transformer and phase-corrector, can be fabricated side-by-side on the same substrate. This solution makes the alignment operation significantly easier, since the number of degrees of freedom is remarkably reduced and the two elements are by-design aligned and parallel to each other. We fabricated compact diffractive optics performing either two-fold or three-fold multiplication and division. The focal length was fixed to 20 mm, therefore the distance between the optics and the mirror was reduced to 1 cm. Adding to the first pattern a spatial frequency around 0.5 µm$^{-1}$, the centres of two phase elements resulted to be around 1 mm from each other. The radius of the first zone was set around 600 µm, while the radii of the phase-correctors were chosen in order not to overlap the first pattern (see Fig. 4 and Fig. 5). In more detail, for both the multipliers we set the parameters *a* and *b* to 300 µm and 250 µm, respectively, while for the dividers we chose the values *a*=250 µm, *b*=400 µm.

**ELECTRON-BEAM LITHOGRAPHY**

The designed optical elements were fabricated as surface-relief phase-only diffractive optics using high-resolution electron-beam lithography (EBL) over a resist layer. By locally controlling the released electronic dose, a different dissolution rate is induced point-by-point on the exposed polymer with a resolution of a few nanometres, giving rise to a spatially-variant resist thickness after the development process [38]. A dose-depth correlation curve (contrast curve) is required to calibrate the correct electron-dose to assign in order to obtain the desired thickness. For a phase pattern $\Omega(x, y)$, the depth $d(x, y)$ of the exposed zone for normal incidence in air is given by:

$$d(x,y) = \frac{\lambda}{n_R(\lambda)-1} \cdot \frac{2\pi - \Omega(x,y)}{2\pi} \qquad (13)$$

being $\lambda$ the incident wavelength, $n_R(\lambda)$ the corresponding refractive index of the resist. In this work, all the diffractive optics have been fabricated by patterning a layer of negative resist (AR-N 7720.30, Allresist), spin-coated on a 1.1 mm thick ITO coated soda lime float glass substrate (resist thickness around 3 µm), and pre-baked for 30 min at 85°C. The fabricated phase-pattern were computed as matrices of pixels with size 0.52 µm and 256 phase levels. At the experimental wavelength of the laser ($\lambda = 632.8$ nm), the refractive index of the resist polymer was assessed to be $n_R = 1.679$ [as measured by spectroscopic ellipsometry (J.A. Woollam VASE, 0.3 nm spectral resolution, 0.005° angular resolution)]. From eq. (13), the maximal depth of the surface relief pattern was found to be 928.5 nm, with a thickness resolution of $\Delta t$ = 3.64 nm. By using custom numerical codes, the phase patterns of the simulated optics were converted into a 3D multilevel structure, which was in turn transformed into a map of electronic doses. A dose correction for compensating the proximity effects was required, in order to both match the layout depth with the fabricated relief and obtain a good shape definition in correspondence of phase discontinuities.

The resist exposure was performed with a JBX-6300FS JEOL EBL machine in high-resolution mode, 12 MHz, generating at 100 KeV and 100 pA an electron-beam with a diameter of 2 nm, providing a resolution down to 5 nm. After exposure, a crosslinking-baking was performed at 100°C for 60 minutes, followed by a post-baking process at 70°C for 2 hours in order to improve surface roughness. Finally, samples were developed for 510 s in AR 300-47 developer (Allresist). After development, the optical elements were gently rinsed in deionized water and blow-dried under nitrogen flux. The quality of the fabricated phase-only optical elements was assessed with scanning electron microscopy (SEM) and optical microscopy. In Fig. 4, SEM inspections are reported, referring to the compact three-fold multiplier. In Fig. 5, inspections refer to the compact three-fold divider.

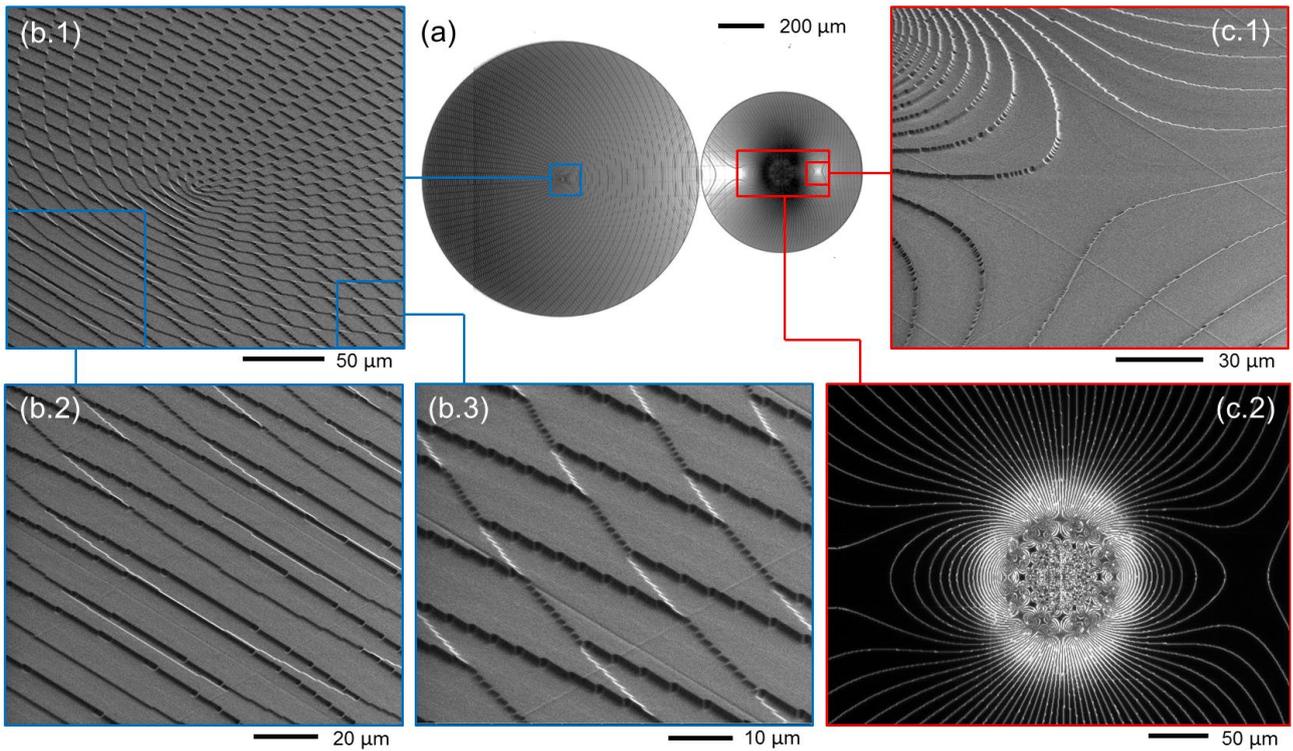

**Figure 4**. Compact two-fold multiplier: inspection at optical microscope (a, c.2 – top view) and SEM analysis (b, c.1 – tilted views). (b.1) SEM inspection of the multiplier central zone and details (b.2, b.3). (c.2) Dark-field optical analysis of the phase-corrector central region, and further details at SEM (c.1).

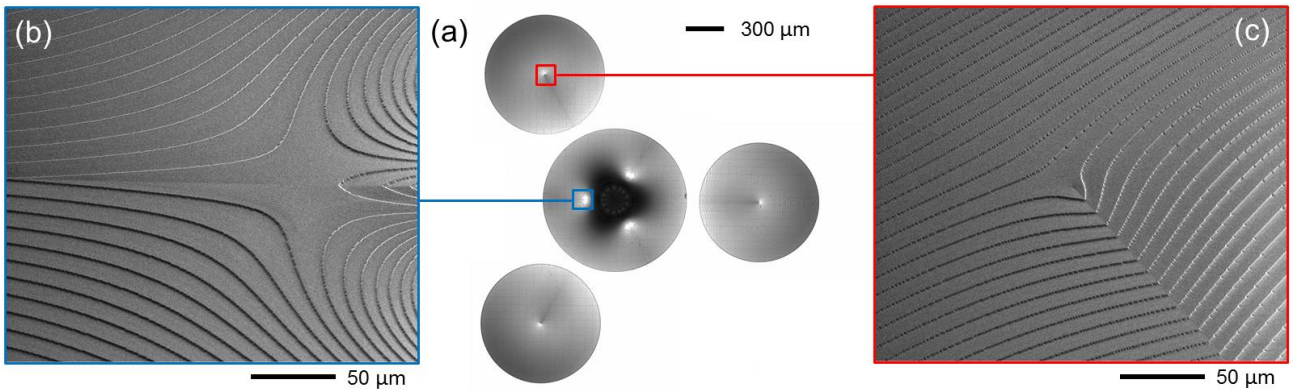

**Figure 5**. Compact three-fold divider: inspection at optical microscope (a, top view) and SEM analysis (b, c – tilted views). (b) SEM inspections of the divider. (c) SEM inspection of one of the three phase-correctors (c).

## OPTICAL CHARACTERIZATION

The experimental layout of the optical characterization bench is shown in Fig. 6a. The optical response of the fabricated optical elements has been analysed for illumination under integer OAM beams, generated with a LCoS spatial light modulator (SLM) (PLUTO-NIR-010-A, Holoeye) using a phase and amplitude modulation technique [39]. The Gaussian beam ($\lambda$= 632.8 nm, beam waist $w_0$=240 μm, power 0.8 mW) emitted by a HeNe laser (HNR008R, Thorlabs) was linearly polarized and expanded with a first telescope ($f_1$=3.5 cm, $f_2$=12.5 cm) before illuminating the SLM display. A second telescope ($f_3$=25.0 cm, $f_4$=12.5 cm) was used in order to isolate and image the first-order encoded mode onto the optical element under test, mounted on a 6-axis kinematic mount (K6XS, Thorlabs). A 50:50 beam-splitter was used to split the beam and check the input beam profile with a first camera (DCC1545M, Thorlabs). The beam illuminated the first zone of the sample, performing either *n*-fold multiplication of division. A mirror was placed on a kinematic mount (KM100, Thorlabs) and its position could be finely controlled with a micrometric translator (TADC-651, Optosigma). The distance from the multiplier/divider was equal to half the focal length of the first element, i.e. 1 cm. After back-reflection, the transformed beam illuminated the optics again of the phase-correcting zones, and the Fourier-transformed beam was collected by a second CMOS camera (DCC1545M, Thorlabs), placed at the back-focal plane of a lens with $f$=10.0 cm. It is worth noting in Fig. 6a that the OAM sign of the two paths is automatically reconciled due to the equal parity in the number of reflections for the beams collected by the two cameras. In order to check the OAM content of the output beams, a Mach-Zehnder interferometric bench was added to the described optical setup, as depicted in Fig. 6a.

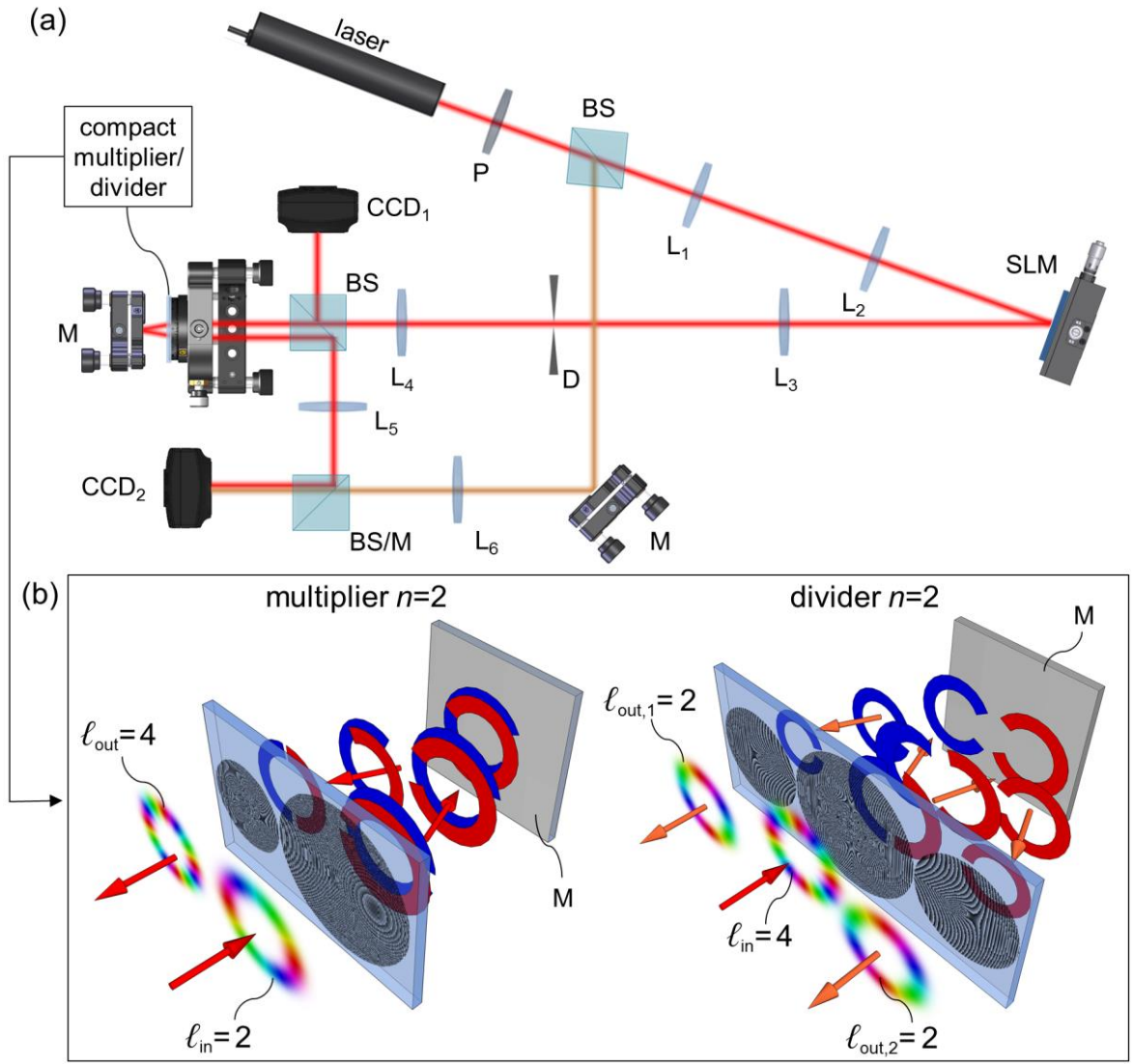

**Figure 6**. (a) Scheme of the experimental setup used for the optical characterization of the fabricated dividers and multipliers. The laser beam is linearly polarized (P) and expanded ($f_1$=3.5 cm, $f_2$=12.5 cm). A beam splitter (BS) is used to generate a reference Gaussian beam for the interferometric bench. The SLM first order is filtered (D) and resized ($f_3$=25.0 cm, $f_4$=12.5 cm) before illuminating the optical element as in (b). A mirror (M) is used for back-reflection on the phase-correcting pattern. A second beam splitter is used both to check the input beam and collect the output on two different cameras. The second camera is placed on the back-focal plane of a fifth Fourier lens ($f_5$=10.0 cm) and it is used to collect the output beam intensity and its interferogram ($f_6$=20.0 cm). (b) Compact optical configurations for the two-fold multiplier and the two-fold divider. The two elements, i.e. tilting divider/multiplier and phase-corrector, are placed on the same substrate and a mirror (M) is used for back-reflection.

## RESULTS AND DISCUSSION

The optical response of the fabricated optics was tested for the multiplication and division of optical beams with integer OAM values. The OAM of the output beam was analysed from the interference pattern with a reference arm obtained by splitting the Gaussian output of the laser. By counting the number of arms in the generated spiralgrams, it is possible to infer the OAM of the output beam and prove the expected multiplication or division of the initial OAM state. The optical characterization of the fabricated samples confirmed the expected capability to perform optical multiplication and division of the OAM of the input beam.

In Fig. 7, the optical characterization is reported for the two-fold multiplier. The optical response was characterized for input OAM beams with $\ell$ in the range from -4 to +4 (Fig. 7a), 0 excluded. As expected, the interference pattern of the output beam shows twice the number of spiral arms of the input interference pattern, confirming the duplication of the input OAM value (Fig. 7c). The same analysis was performed for the three-fold multiplier, for input OAM values in the range from -3 to +3, 0 excluded (Fig. 8a). As shown in Fig. (8c), the number of output arms in the interferograms is three times the input one, as expected.

In Fig. 9, the experimental output of the two fold-divider is reported, for input OAM beams with even $\ell$ in the range from -8 to +8, 0 excluded (Fig. 9a). As expected, the input beam is split into two output beams (Fig. 9b), with OAM equal to half the input value. In Fig. 9c, the interferogram is reported for only one out of the two beams, and the interference pattern confirms the capability of the designed optics to divide the input OAM into two. The same analysis was performed for the three-fold divider, as reported in Fig. 10. In this case, the output is constituted of three OAM beams (Fig. 10b), each one carrying one third of the input OAM, as demonstrated by the experimental interferograms (Fig. 10c) for input $\ell$ in the range from -9 to +9, with a step of 3, 0 excluded (Fig. 10a).

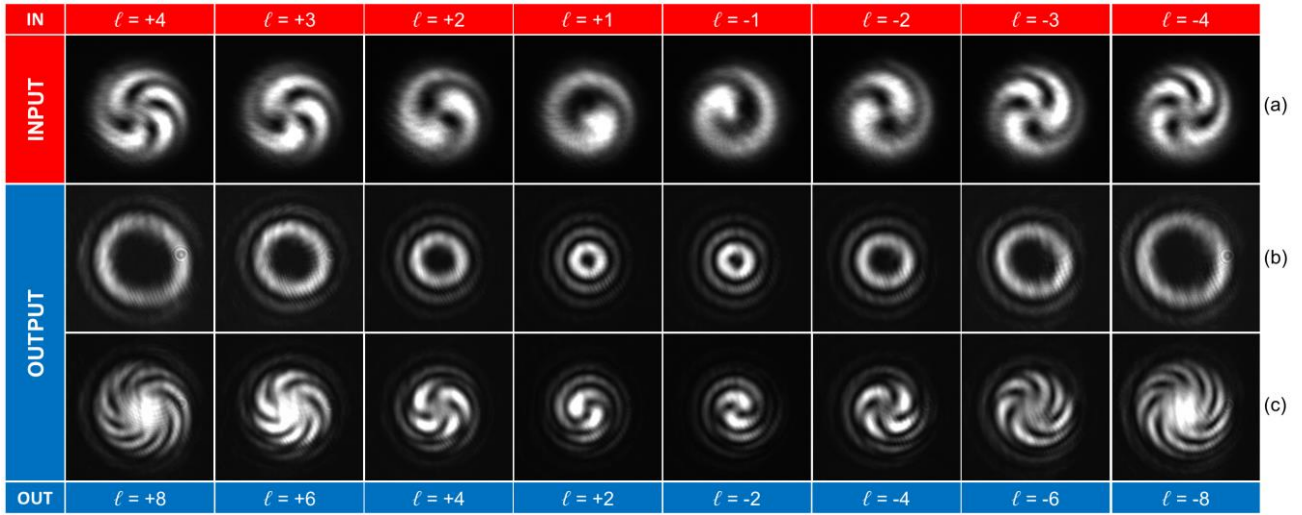

**Figure 7**. Optical characterization of the two-fold multiplier for input OAM beams with $\ell$ in the range from -4 to +4, 0 excluded. (a) Interferogram of the input OAM beams. Intensity pattern (b) and interferogram (c) of the output beam. As expected, the number of spiral arms in the output interference pattern is equal to twice the number in the input interferogram.

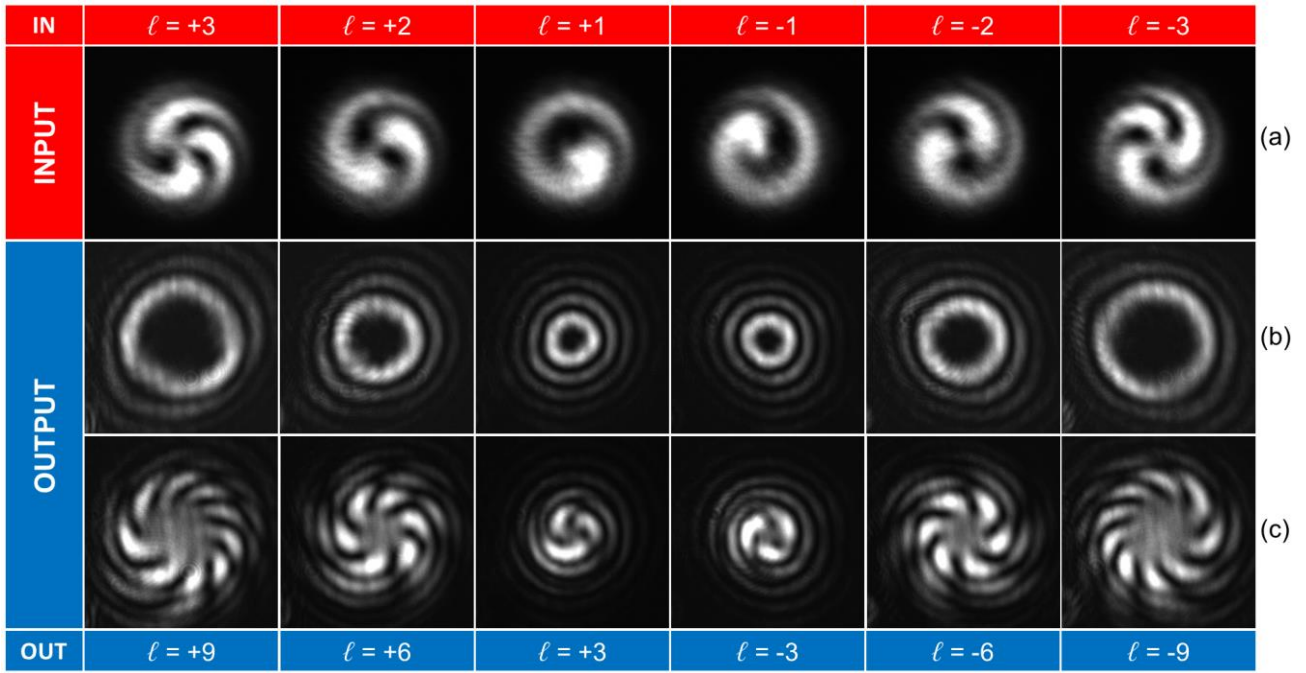

**Figure 8**. Optical characterization of the three-fold multiplier for input OAM beams with $\ell$ in the range from -3 to +3, 0 excluded. (a) Interferogram of the input OAM beams. Intensity pattern (b) and interferogram (c) of the output beam. As expected, the number of spiral arms in the output interference pattern is equal to three times the number in the input interferogram.

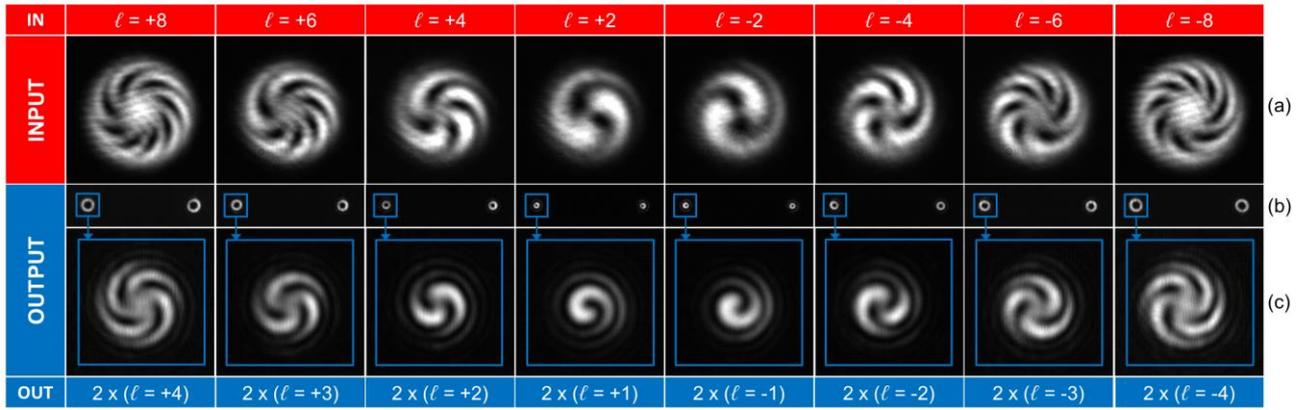

**Figure 9**. Optical characterization of the two-fold divider for input OAM beams with even $\ell$ in the range from -8 to +8, 0 excluded. (a) Interferogram of the input OAM beams. Intensity pattern (b) and interferogram (c) of the output beam. As expected, the number of spiral arms in the output interference pattern is equal to half the number in the input interferogram.

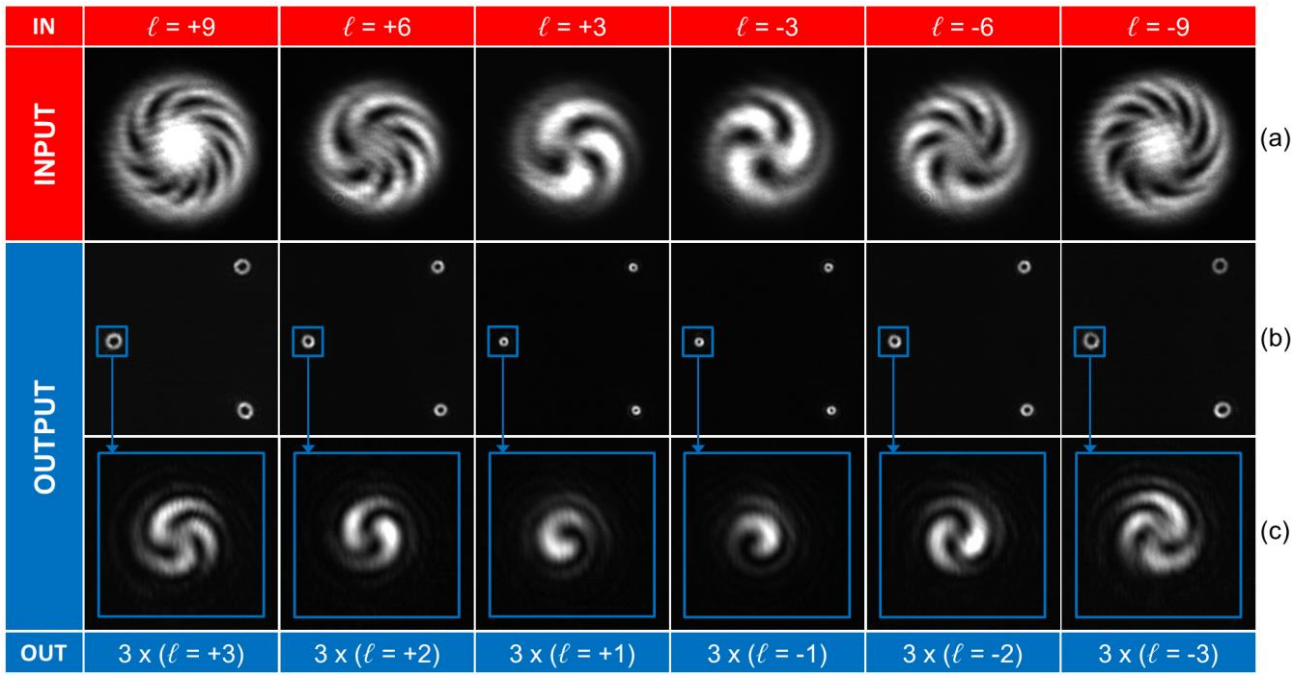

**Figure 10**. Optical characterization of the three-fold divider for input OAM beams with $\ell$ in the range from -9 to +9, with a step of 3, 0 excluded. (a) Interferogram of the input OAM beams. Intensity pattern (b) and interferogram (c) of the output beam. As expected, the number of spiral arms in the output interference pattern is equal to one third the number in the input interferogram.

# CONCLUSIONS

In this paper, we have presented for the first time a comprehensive work of simulation, fabrication and optical characterization of a new type of optical elements performing the division and multiplication of the orbital angular momentum of light. With respect to previous demonstrations, these optical operations are here performed in a compact and efficient configuration by means of a cascade of only two confocal optical elements. The former performs the desired optical transformation, i.e. $n$-fold multiplication or division of the input beam, while the latter introduces the required phase-correction in order to consider the distortion accumulated during beam propagation. Both multiplication and division are implemented with a superposition of many circular-sector transformations. In the specific, multiplication by a factor $n$ is realized by splitting the input OAM beam into $n$ copies and mapping the whole azimuthal gradients into $n$ complementary circular sectors, thus obtaining a final azimuthal gradient which is $n$-times the input one. Conversely, OAM division is achieved by splitting the input azimuthal gradient into $n$ complementary circular-sectors and mapping them onto $n$ complete circles at distinct positions.

The phase-patterns of the phase-only optical elements performing two-fold and three-fold division/multiplication have been calculated in the paraxial approximation, and implemented in the form of diffractive optics fabricated with high-resolution electron-beam lithography. Furthermore, the compactness of the optical system has been further improved by fabricating the two optical elements, i.e. multiplier/divider and corresponding phase-corrector, side-by-side onto the same substrate, therefore reducing the degrees of freedom and significantly improving alignment and miniaturization. This is achievable by introducing a tilt in the first optical element and using a mirror for back-reflection onto the second pattern performing phase-correction. The optical characterization with input integer OAM values confirmed the expected capability of the fabricated samples to perform either multiplication or division of the OAM content of the input beams.

So far, the possibility to increase and decrease the OAM content of a beam in a multiplicative manner was still missing, therefore the framework offered by these optical elements can open to interesting and promising applications in a wide range of fields, both in the classical and single-photon regimes. OAM multiplication could find applications in the generation of high-order OAM states, for structured-light studies, or for incremental particle tweezing and trapping/manipulation. By using many low-order multipliers in cascade, it could be possible to easily increase the OAM of one order of magnitude or more. By increasing the input OAM using a two-fold or three-fold multiplier before a standard *log-pol* sorter, it could be possible to significantly enhance the cross-talk by improving the channel separation without the need of a fan-out optical operation. Multipliers and dividers could represent the key elements to implement OAM algebra in optical boards performing computation with the OAM of light. Moreover, they can allow the compact and efficient implementation of the operations required in next-generation optical platforms performing switching and routing between OAM channels in mode-division multiplexing architectures. In this paper, OAM division was limited to an equalized set of output modes, however it could be easily generalized for the generation of any complementary set of either integer or fractional OAM states.

The theory underlying the division and multiplication of azimuthal phase gradients, peculiar of modes endowed with orbital angular momentum, can be extended beyond the specific optical range of the presented study. For instance, the exploitation of mode-division multiplexing for high-capacity transmissions in free-space has acquired increasing attention in the microwave regime, and the implementation of diffractive optics for OAM control and manipulation has been recently shown [40]. Moreover, the interest for physical states carrying OAM has recently encompassed other branches of physics, by applying the results of optics to quantum mechanical wave functions describing massive particles (matter waves) [41]. In particular, the possibility to control and analyze the OAM of free-electron beams has been recently investigated with diffractive [42, 43] or electrostatic phase [44] elements. Therefore, we envisage that similar devices performing multiplication/division could be developed for studies concerning the multiplicative transformation of the OAM carried by twisted matter waves.

## ACKNOWNLEDGEMENTS

This work was supported by project Vortex 2 from CEPOLISPE.

## AUTHOR CONTRIBUTIONS

G.R. conceived the idea, performed the design and optimization of the phase pattern and conducted optical characterization. M.M. performed EBL fabrication, and optical and SEM microscopy. F.R. addressed the aims of the project and managed the laboratory. All authors discussed and contributed to the writing of the manuscript.

## CONFLICTS OF INTEREST

The authors declare no competing financial interests.